# A reversible high embedding capacity data hiding technique for hiding secret data in images

Mr. P. Mohan Kumar,
Asst. Professor, CSE Department,
Jeppiaar Engineering College,
Chennai., India.
mohankumarmohan@gmail.com

Dr. K. L. Shunmuganathan,
Professor and Head, CSE Department
R.M.K. Engineering College,
Chennai. India.
kls_nathan@yahoo.com

*Abstract --* **As the multimedia and internet technologies are growing fast, the transmission of digital media plays an important role in communication. The various digital media like audio, video and images are being transferred through internet. There are a lot of threats for the digital data that are transferred through internet. Also, a number of security techniques have been employed to protect the data that is transferred through internet. This paper proposes a new technique for sending secret messages securely, using steganographic technique. Since the proposed system uses multiple level of security for data hiding, where the data is hidden in an image file and the stego file is again concealed in another image. Previously, the secret message is being encrypted with the encryption algorithm which ensures the achievement of high security enabled data transfer through internet.**

## I. INTRODUCTION

Steganography is the technique of hiding information. The primary goal of cryptography is to make a data that cannot be understood by a third party, where as the goal of steganography is to hide the data from a third party. There are many number of steganographic methods ranging from invisible ink and microdots to hide a secret message in the second letter of each word of a large body of text and spread spectrum radio communication. With the vast development of computers and internet, there are many other methods of hiding information [1], such as:

   a. Covert channels
   b. Concealment of text message within Web pages
   c. Hiding files in "plain sight"
   d. Null ciphers

One of the most important applications of steganography is digital watermarking. A watermark is the replication of an image, logo, or text on paper stock so that the source of the document can be at least partially authenticated. A digital watermark can accomplish the same function; an artist can post sample images on his website with an embedded signature so that he can prove her ownership in case others attempt to steal his work or try to show as their work.

The following formula can provide a very generic description of the steganographic process:
Cover data + hidden data + stego key = stego data

In this formula, the cover data is the file in which we will hide the hidden data, which may also be encrypted using the stego key. The resultant file is the stego data which will be of the same type as the cover data [2]. The cover data and stego data are typically image or audio files. In this paper, we are going to focus on image files and will discuss about the existing techniques of image steganography.

Before discussing how information is hidden in an image file, we should have an idea about how images are stored. An image file is simply a binary file containing a binary representation of the color or light intensity of each picture element known as pixel, comprising the image.

Images are normally using either 8-bit or 24-bit color. When using 8-bit color, there is a definition of up to 256 colors forming a palette for this image, where each color is denoted by an 8-bit value. A 24-bit color scheme uses 24 bits per pixel which provides a much better set of colours. In this case, each pixel is represented by three bytes, each byte representing the intensity of the three primary colors red, green, and blue (RGB), respectively[3].

The size of an image file is directly related to the number of pixels and the granularity of the color definition. A typical 640x480 pix image using a palette of 256 colors would require a file about 307 KB in size (640 • 480 bytes), whereas a 1024x768 pix high-resolution 24-bit color image would result in a 2.36 MB file (1024 • 768 • 3 bytes).

There are a number of image compression schemes have been developed as Bitmap (BMP), Graphic Interchange Format (GIF), and Joint Photographic Experts

Group (JPEG) file types. Anyway, we are not able to use them all as the same way for steganography.

GIF and 8-bit BMP files are using lossless compression, a scheme that allows the software to exactly reconstruct the original image. JPEG, on the other hand, uses lossy compression, which means that the expanded image is very nearly the same as the original but not an exact duplicate. While both of these methods allow computers to save storage space, lossless compression is much better suited to applications where the integrity of the original information must be maintained, such as steganography. Even though JPEG can be used for stego applications, more commonly used files for hiding data are GIF or BMP files.

## II. LITERATURE SURVEY

The rapid advances of network technologies and digital devices make information exchange fast and easy. However, distributing digital data over public networks such as the Internet is not really secure due to copy violation, counterfeiting, forgery, and fraud. Therefore, protective methods for digital data, specially for sensitive data, are highly demanded. Traditionally, secret data can be protected by cryptographic methods such as DES and RSA (Rivest et al., 1978) [4]. The drawback of cryptography is that cryptography can protect secret data in transit, but once they have been decrypted, the content of the secret data has no further protection (Cox et al., 2007).

In addition, cryptographic methods do not hide the very existence of the secret data. Alternatively, confidential data can be protected by using information hiding techniques. Information hiding embeds secret information into cover objects such as written texts, digital images, adios, and videos (Bender et al., 1996) [5]. For more secure, cryptographic techniques can be applied to an information hiding scheme to encrypt the secret data prior to embedding.

In general, information hiding (also called data hiding or data embedding) technique includes digital watermarking and steganography (Petitcolas et al., 1999). Watermarking is used for copyright protection, broadcast monitoring, transaction tracking, etc. A watermarking scheme imperceptibly alters a cover object to embed a message about the cover object (e.g., owner's identifier) (Cox et al., 2007). The robustness (i.e. the ability to resist certain malicious attacks such as common signal processing operations) of digital watermarking schemes is critical. In contrast, steganography is used for secret communications.

A steganographic method undetectably alters a cover object to embed a secret message (Cox et al., 2007) [6]. Thus, steganographic methods can hide the very presence of covert communications. Information hiding techniques can be performed in three domains (Bender et al., 1996) [7], namely, spatial domain (Zhang and Wang, 2006), compressed domain (Pan et al., 2004), and frequency (or transformed) domain (Kamstra and Heijmans, 2005; Wu and Frank, 2007; Zhou et al., 2007) [8].

Each domain has its own advantages and disadvantages in terms of embedding capacity, execution time, storage space, etc. Two main factors that really affect an information hiding scheme are visual quality of stego images (also called visual quality for short), embedding capacity (or payload). An information hiding scheme with low image distortion is more secure than that with high distortion because it does not raise any suspicions of adversaries. The second important factor is embedding capacity (also called capacity for short).

An information hiding scheme with high payload is preferred because more secret data can be transferred [9]. However, embedding capacity is inversely proportional to visual quality. Thus, the tradeoff between the two factors above varies from application to application, depending on users' requirements and application fields. Consequently, different techniques are utilized for different applications. Therefore, a class of data hiding schemes is needed to span the range of possible applications. Embedding the secret data into an image causes the degradation of image quality. Even though small image distortion is unacceptable in some applications such as law enforcement, military image systems, and medical diagnosis.

If a data embedding scheme is irreversible (also called lossy), then a decoder can extract secret data only and the original cover image cannot be restored. In contrast, a reversible (also called invertible, lossless, or distortion-free) data embedding scheme allows a decoder to recover the original cover image completely upon the extraction of the embedded secret data [10]. A reversible data hiding scheme is suitably used for some applications such as the healthcare industry and online content distribution systems.

To our best knowledge, the first reversible data embedding scheme was proposed in 1997 (Barton, 1997). Macq (2000) extended the patchwork algorithm (Bender et al., 1996) [11] to achieve the reversibility. This method encounters the underflow and overflow problem (i.e., grayscale pixel values are out of the allowable range [0, 255]). Honsinger et al. (2001) [12] used modulo arithmetic operation to resolve the underflow and overflow problem.

Consequently, Honsinger et al.'s method raises the salt-and-pepper effect. Fridrich et al. (2001) [13] also proposed the reversible data embedding method for the authentication purpose so the embedding capacity of this method is low. Later on, De Vleeschouwer et al. (2003) [14] proposed the circular interpretation of bijective transforms to face the underflow and overflow problem. However, the salt-and-pepper problem still remains in De Vleeschouwer et al.'s method.

As a whole, the problem with the aforementioned methods is either the salt-and-pepper problem or low embedding capacity. Tian (2003) [15] proposed the reversible data embedding scheme with high embedding capacity and good visual quality of embedded images (also called stego images). Tian's scheme is of a fragile technique meaning that the embedded data will be mostly destroyed when some common signal processing operations (e.g., JPEG compression) are applied to a stego image. Tian's method uses the difference expansion (DE) operation to hide one secret bit into the difference value of two neighboring pixels. Thus, the embedding capacity of the DE method is at most 0.5 bpp for one layer embedding. Tian also suggested the multiple-layer embedding to achieve higher embedding capacity. Alattar (2004) [16] generalized Tian's method to embed n _ 1 secret bits into a group of n cover pixels. Thus, the embedding capacity of Alattar's method is at most (n _ 1)/n bpp.

Kamstra and Heijmans (2005) [17] also improved Tian's method in terms of visual quality at low embedding capacities. The maximum embedding capacity of Kamstra and Heijmans' method is 0.5 bpp. Chang and Lu (2006) exploited Tian's method to achieve the average embedding capacity of 0.92 bpp and the average PSNR of 36.34 dB for one-layer embedding. Next, Thodi and Rodriquez (2007) improved Tian's scheme and proposed the novel method called prediction error expansion (PEE) embedding. The PEE method embeds one secret bit into one cover pixel at a time. However, at its maximum embedding capacity (i.e., around 1 bpp), the visual quality of the PEE method is always less than 35 dB for all test images. Then, Kim et al. (2008) improved Tian's method by simplifying the location map to achieve higher embedding capacity while keeping the image distortion the same as the original DE method. Lou et al. (2009) improved the DE method by proposing the multiple layer data hiding scheme. Lou et al.'s method reduces the difference value of two neighboring cover pixels to enhance the visual quality. The problem with the aforementioned schemes is that the PSNR value becomes very low (i.e., less than 30 dB) at high embedding capacity (i.e., more than 1 bpp).

III. PROPOSED SYSTEM

This section presents our new reversible steganographic scheme with good stego-image quality and high payload by using the multiple embedding strategies to improve the image quality and the embedding capacity of the DE method. For increasing the security of secret data delivery, it is assumed that the secret data have been encrypted by using the well-known cryptosystem (e.g., DES or RSA) to encrypt the secret data prior to embedding. Therefore, even an attacker somehow extracts the secret data from the stego image; the attacker still cannot obtain the real information without the decryption key. The details of the proposed method are described next.

*A. The embedding phase*

Basically, the proposed method embeds one information bit b of the information bit stream into one grayscale cover pixel pair of an original grayscale cover image O sized H _W at a time in raster scan order. Specifically, the proposed scheme consists of two main stages, namely, the horizontal embedding procedure HEm and the vertical embedding procedure VEm. The secret bit stream S whose length is LS is divided into two secret bit streams S1 and S2. The lengths of S1 and S2 are denoted as LS1 and LS2, respectively. The information bit stream B1 is created by concatenating the secret bit stream S1 and the auxiliary data bit stream A1. That is, B1 = S1||A1.

Similarly, the information bit stream B2 is created by concatenating the secret bit stream S2 and the auxiliary data bit stream A2 (i.e., B2 = S2||A2). The generation of A1 and A2 will be described later. Firstly, the information bit stream B1 is horizontally embedded into O by using the procedure HEm to obtain the output image T sized H _W. Secondly, the compressed location map CM1 whose length is LC1, which will be described later, is embedded into T by using the least significant bit (LSB) replacement technique to obtain the output image U sized H _W. Thirdly, the information bit stream B2 is vertically embedded into U by using the procedure VEm to obtain the output image V sized H _W. Fourthly, the compressed location map CM2 whose length is LC2, which will be described later, is embedded into V by using the LSB replacement technique to obtain the final stego image X sized H _ W.

The overview of the proposed embedding process is shown in the following diagram. For the horizontal embedding procedure HEm: horizontally scan the cover image O in raster scan order (i.e., from left to right and top to bottom) to gather two neighboring pixels x and y into a cover pixel pair (x, y). If y is an odd value, then the cover pixel pair (x, y) is defined as a horizontally embeddable pixel pair. Otherwise, the cover pixel pair (x, y) is defined as a horizontally non-embeddable pixel pair. Let the set of horizontally embeddable pixel pairs of O be E1 whose cardinality is LE1. It is clear that the length of B1 is LE1. The horizontally non-embeddable pixel pairs are kept unchanged during the horizontal embedding stage. Each information bit b in B1 is horizontally embedded into each horizontally embeddable pixel pair (x, y) in E1 at a time by using the proposed horizontal embedding rule HR defined below.

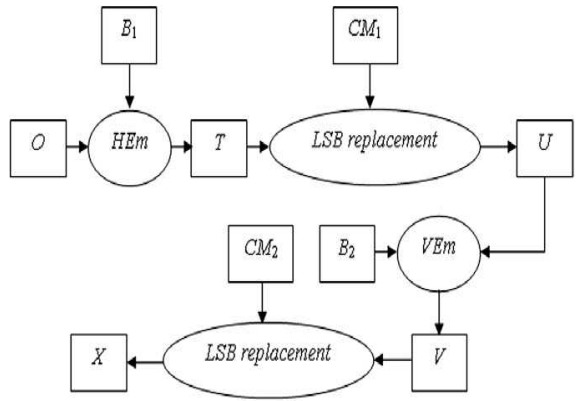

Fig 1. Embedding Phase of Proposed system

*The horizontal embedding rule HR:*

For each horizontally embeddable pixel pair (x, y), we apply the following embedding rules:

HR1: If the information bit b = 1, then the stego pixel pair is computed by (x0, y0) = (x, y).

HR2: If the information bit b = 0, then the stego pixel pair is calculated by (x0, y0) = (x, y _ 1).

The horizontal embedding rule HR is repeatedly applied to embed each information bit b in B1 into each cover pixel pair (x, y) in E1 of O until the whole information bit stream B1 is completely embedded into O to obtain the output image T. It is noted that the proposed horizontal embedding rule HR does not cause the underflow and overflow problem. That is, the embedded pixel pairs (x0, y0)'s are guaranteed to fall in the allowable range [0, 255].

The auxiliary data bit stream A1 is actually the LSBs of the first LC1 pixels in the image T and generated as follows. It is noted that LC1 is the length of the compressed location map CM1 ended with the unique end-of-map indicator EOM1. Initially, B1 is equal to S1 (i.e., B1 = S1). During the execution of the procedure HEm, for the first LC1 pixels in O, when each pixel has been processed for embedding, its LSB is taken as an auxiliary data bit of A1 and appended to the end of B1. That is, B1 is gradually grown until the LC1 auxiliary data bits in A1 are concatenated into B1. Finally, the information bit stream is B1 = S1||A1, which is completely embedded into O.

*For the vertical embedding procedure VEm:*

Vertically scan the output image U in raster scan order to group two neighboring pixels u and v into a pixel pair (u, v). If v is an even value, then the pixel pair (u, v) is defined as a vertically embeddable pixel pair. Otherwise, the pixel pair (u, v) is defined as a vertically non-embeddable pixel pair. Let the set of vertically embeddable pixel pairs of U be E2 whose cardinality is LE2. It is obvious that the length of B2 is LE2. The vertically non-embeddable pixel pairs are left unchanged during the vertical embedding stage. Each information bit b in B2 is vertically embedded into each vertically embeddable pixel pair (u, v) in E2 at a time by using the proposed vertical embedding rule VR defined below.

*The vertical embedding rule VR:*

For each vertically embeddable pixel pair (u, v), we apply the following embedding rules:

VR1: If the information bit b = 0, then the final stego pixel pair is computed by (u0, v0) = (u, v).

VR2: If the information bit b = 1, then the final stego pixel pair is computed by (u0, v0) = (u, v + 1).

The vertical embedding rule VR is iteratively applied to conceal each information bit b in B2 into each pixel pair (u, v) in E2 of U until the entire information bit stream B2 is totally concealed into U to obtain the output image V. It is noted that the proposed vertical embedding rule VR does not raise the underflow and overflow problem. That is, the final stego pixel pairs (u0, v0)'s are assured to fall in the allowable range [0, 255]. Similar to the generation of A1, the auxiliary data bit stream A2 is actually the LSBs of the first LC2 pixels in the image V and generated as follows. It is noted that LC2 is the length of the compressed location map CM2 ended with the unique end-of-map indicator EOM2.

Initially, B2 equals the secret bit stream S2 (i.e., B2 = S2). During the execution of the procedure VEm, for the first LC2 pixels in the image U, when each pixel has been processed for embedding, its LSB is taken as an auxiliary data bit of A2 and appended to the end of B2.

That is, B2 is gradually grown until the LC2 auxiliary data bits in A2 are concatenated into B2. Finally, the information bit stream is B2 = S2||A2, which is fully embedded into the image U. For the purposes of extracting B1 and recovering O, a location map HL sized H _ (W/2) is needed to record the positions of the horizontally embeddable pixel pairs (x, y) in O. The location map HL is a one-bit bitmap.

All the entries of HL are initialized to 0. If the cover pixel pair (x, y) is the horizontally embeddable pixel pair, then the corresponding entry of HL is set to be 1. Next, the location map HL is losslessly compressed by using the JBIG2 codec (Howard et al., 1998) or an arithmetic coding toolkit (Carpenter, 2002) to obtain the compressed location map CM1 whose length is LC1. The compressed location map CM1 is embedded into the image T by using the LSB replacement technique as mentioned above. Similarly, for the purposes of extracting B2 and recovering the image U, a location map VL sized (H/2) _W is required to save the positions of the vertically embeddable pixel pairs (u, v) in U. The location map VL is a one-bit bitmap.

All the entries of VL are initialized to 0. If the pixel pair (u, v) is the vertically embeddable pixel pair, then the corresponding entry of VL is set to be 1. Then, VL is also lossless compressed by using the JBIG2 codec (Howard et al., 1998) or an arithmetic coding toolkit (Carpenter, 2002) to obtain the compressed location map CM2 whose length is LC2. The compressed location map CM2 is embedded into the image V by using the LSB replacement technique as mentioned above. The final output of the embedding phase is the final stego image X sized H _W. Then, the stego image X is sent to the expected receivers.

### B. The extracting phase

The extracting phase is actually the reverse process of the embedding phase. The extracting phase is composed of two main stages, namely, the vertical extracting procedure VEx and the horizontal extracting procedure HEx. Specifically, firstly, the embedded CM2 is retrieved by extracting the LSBs of the first LC2 pixels of the received stego image X. The extracted CM2 is then decompressed to obtain VL which is used to identify the vertical embeddable pixel pairs belonging to the set E2 of X. Next, A2 is extracted from the last LC2 pixel pairs in E2 of X by using the vertical extracting rule VX. Then, the first LC2 pixel pairs of X are replaced with the extracted A2 to obtain the image V. Secondly, from the image V, extract the embedded B2 and recover the image U by using the vertical extracting procedure VEx. Thirdly, the embedded CM1 is obtained by extracting the LSBs of the first LC1 pixels of the image U. The extracted CM1 is then decompressed to obtain HL which is used to identify the horizontal embeddable pixel pairs belonging to the set E1 of U.

Next, A1 is extracted from the last LC1 pixel pairs in E1 of U by using the horizontal extracting rule HX. Then, the first LC1 pixel pairs of U are replaced with the extracted A1 to obtain the image T. Fourthly, from the image T, extract the embedded B1 and recover the original cover image O by using the horizontal extracting procedure HEx. The first LS1 bits of B1 is the secret bit stream S1 and the first LS2 bits of B2 is the secret bit stream S2. The extracted secret bit streams S1 and S2 are concatenated to form the original secret bit stream S (i.e., S = S1||S2.). The overview of the proposed extracting process is shown in the following figure.

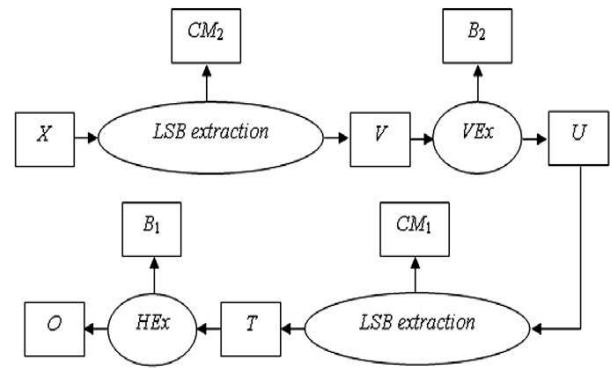

Fig.2. Extracting phase of proposed system

*For vertical extracting procedure VEx*

Vertically scan the image V in raster scan order to group two neighboring pixels u0 and v0 into a pixel pair (u0 , v0). The extracted VL is used to determine whether a pixel pair (u0 , v0) belongs to the set E2 (i.e., a vertically embeddable pixel pair). The extraction of the embedded B2 and the recovery of the image U are performed as follows.

The vertical extracting rule VX

If v0 is an even value,

then The information bit in B2 is extracted by b = 0 and The pixel pair (u, v) is recovered by (u, v) = (u0 , v0).

Else if (u0 , v0) belongs to the set E2,

thenThe information bit in B2 is extracted by b = 1 and

The pixel pair (u, v) is recovered by (u, v) = (u0 , v0 _ 1).

Else

There is no information bit extraction and

The pixel pair (u, v) is recovered by (u, v) = (u0, v0).

The output of the vertical extracting procedure VEx is the image U.

From the image U, the embedded CM1 is extracted and the image T is recovered as mentioned above.

The location map HL is achieved from decompressing the extracted CM1.

*For horizontal extracting procedure VEx*

Horizontally scan the image T in raster scan order to gather two neighboring pixels x0 and y0 into a pixel pair (x0 , y0). The location map HL is used to identify if a pixel pair (x0 , y0) belongs to the set

E1 (i.e., a horizontally embeddable pixel pair). The extraction of the embedded B1 and the recovery of the original cover image O are performed as below.

The horizontal extracting rule HX

If y0 is an odd value, then

The information bit in B1 is extracted by b = 1 and

The original cover pixel pair (x, y) is recovered by (x, y) = (x , y).

Else if (x0 , y0) belongs to the set E1, then

The information bit in B1 is extracted by b = 0 and

The original cover pixel pair (x, y) is recovered by (x, y) = (x0 , y0 + 1).

Else

There is no information bit extraction and

The original cover pixel pair (x, y) is recovered by (x, y) = (x0 , y0).

## IV. EXPERIMENTAL RESULTS

To evaluate the performance of the proposed method, we implemented the proposed method and Tian's method by using Borland C++ Builder 6.0 software running on the Pentium IV, 3.6 GHz CPU, and 1.49 GB RAM hardware platform. The secret bit stream S was randomly generated by using the library function random(). The multiple-layer embedding was performed for the DE and proposed methods. To make the DE method achieve its maximum embedding capacity, the threshold TH was not used in the experiments. The location maps L, HL, and VL were losslessly compressed and decompressed by using the arithmetic coding toolkit (Carpenter, 2002). The commonly used grayscale images sized 512 _ 512, were used as the cover images in our experiments. The good visual quality of stego images (i.e. images embedded with a secret message) is the most important property of steganographic systems because it is hard to be detected by detectors. Because the lack of a universal image quality measurement tool, we used peak signal-to-noise ratio (PSNR) to measure the distortion between an original cover image and the stego image. The PSNR is defined by

$$PSNR = 10\log_{10}\frac{255^2}{MSE} \text{ (dB)}, \text{ where}$$

$$MSE = \frac{1}{H \times W}\sum_{i=1}^{H}\sum_{j=1}^{W}(O_{ij} - X_{ij})^2,$$

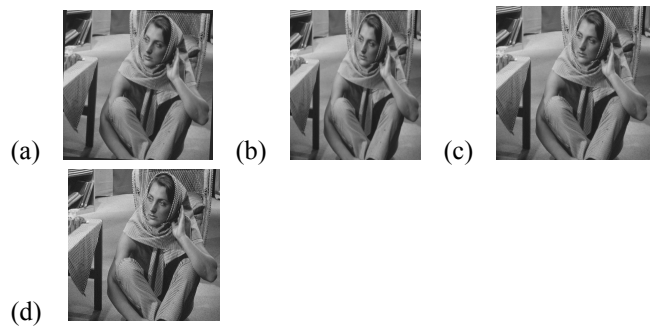

(a)  (b)  (c)

(d)

Fig 3.a. Host image b. Image after preprocessing c. Stego image d. Image quality after extracting secret image

## V. CONCLUSION

In this paper, we propose a simple reversible steganographic scheme in spatial domain for digital images by using the proposed multiple embedding strategies. The experimental results show that the proposed reversible steganographic method is capable of achieving very good visual quality of stego images and high embedding capacity (especially, when multiple-layer embedding is performed). Specifically, with the one-layer embedding, the proposed method can obtain the embedding capacity of more than 0.5 bpp and the PSNR value greater than 54 dB for all test images. In addition, with the two-layer embedding, the proposed method can achieve the embedding capacity of about 1 bpp and the PSNR value greater than 53 dB for all test images. Especially, with the five-layer embedding, the proposed method has the embedding capacity of more than 2 bpp and the PSNR value higher than 52 dB for all test images. Therefore, it can be said that the proposed method is the one that really allows users to perform multiple layer embedding to achieve the purposes of very high embedding capacity and very good visual quality of stego images. As a whole, the proposed method outperforms many existing reversible data embedding methods in terms of visual quality, embedding capacity, and computational complexity. Thus, we can conclude that our proposed method is applicable to some information hiding applications such as secret communications, medical imaging systems, and online content distribution systems.

## ACKNOWLEDGEMENT

We take immense pleasure in thanking our chairman Dr. Jeppiaar M.A, B.L, Ph.D, the Directors of Jeppiaar Engineering College Mr. Marie Wilson, B.Tech, MBA, (Ph.D), Mrs. Regeena Wilson, B.Tech, MBA, (Ph.D) and the principal Dr. Sushil Lal Das M.Sc(Engg.), Ph.D for their continual support and guidance. We would like to extend our thanks to my guide, our friends and family members without whose inspiration and support our efforts would not have come to true. Above all, we would like to thank God for making all our efforts success.

## AUTRHORS PROFILE

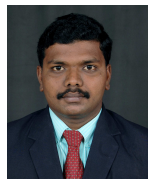

P. Mohan Kumar B.E.,M.E.,(Ph.D) works as Assistant Professor in Jeppiaar Engineering College and he has more than 8 years of teaching experience. His areas of specializations are Network security, Image processing and artificial intelligence.

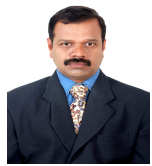

Dr. K.L. Shanmuganathan B.E, M.E.,M.S.,Ph.D works as the Professor & Head of CSE Department of RMK Engineering College, Chennai, TamilNadu, India. He has more than 18 years of teaching experience and his areas of specializations are Artificial Intelligence, Computer Networks and DBMS.